# The effect of surface preparation on high temperature oxidation of Ni, Cu and Ni-Cu alloy

Daria Serafin*[1], Wojciech J. Nowak[1], Bartek Wierzba[1]

[1] Department of Materials Science, Faculty of Mechanical Engineering and Aeronautics, Rzeszow University of Technology, al. Powstancòw Warszawy 12, 35-959 Rzeszów, Poland

*corresponding author: Daria Serafin, e-mail address: darseraf@gmail.com, phone: +48 665 772 816

**Abstract**

In the present paper, the analysis of polishing, grinding and sand-blasting processes on oxidation behavior of nickel, copper and Ni30Cu70 alloy has been performed for three different temperatures. Surface roughness has been analyzed using two methods: contact profilometer and fractal analysis. It has been proved that Ra parameter for sand-blasted samples is greater about one order of magnitude from ground and two order of magnitude from polished samples. Obtained mass gain plots after isothermal oxidation tests, SEM microphotographs and SEM/EDS analysis of the samples prove that morphology of the polished and ground samples does not vary significantly after oxidation at investigated temperatures, but differences in oxide layer thicknesses and mass gains were observed. Oxidized sand-blasted samples revealed that increased surface roughness and alumina intrusions remained in the surface after sand-blasting process change the oxidation process– mixed Al/(Ni, Cu or Ni-Cu) oxide is formed and separate $Al_2O_3$ particles are visible. Moreover, differences in thickness ratio of copper oxides: $Cu_2O$ to $CuO$ on copper samples oxidized at 750 °C (26, 20; 29, 64; 43, 45 for polished, ground and sandblasted samples, respectively) prove that surface preparation influence high-temperature oxidation of the samples by promoting new diffusion mechanism.

# 1. Introduction

Oxidation of pure metals has been studied over the years by many authors [1,2]. Copper and nickel make no exception. The oxidation of the former was performed in wide temperature range by Mrowec and Stoklosa [3] and Zhu et al. [4,5] because copper and its alloys are commonly used as an engineering materials due to their high ductility, electrical and thermal conductivity and quite good strength [6]. It has been established, that at high temperature range (600–900 °C) oxidation of copper obeys the parabolic rate law and results in formation of two oxides: $Cu_2O$ and $CuO$. The main oxidation product however – $Cu_2O$, grows as a result of outward diffusion of copper – by simultaneous mechanisms of lattice and grain boundary diffusion [3,4,7,8].

Also oxidation of nickel has been investigated, mainly because it forms only NiO during exposure at high temperature, what made it relatively easy to use it as a basis of theoretical calculations, which improved understanding of oxidation processes [1,2]. Even though NiO scales form over a wide temperature range, it is known that only at temperatures higher than 900 °C, the lattice diffusion starts to prevail over grain boundary diffusion and the oxidation rate is parabolic [2,9]. At lower temperatures, that is between 300-450°C the oxidation process follows quadratic dependence on time [10]. At intermediate temperature range: 700-900°C, the oxidation rate is initially rapid then decreases to parabolic rate or continuously decreases further [9, 10].

Nickel forms solid solution with copper at a whole range of concentration. Addition of copper dramatically improves alloy's thermal conductivity and its resistance to seawater at low temperature [11]. From the other side nickel is alloyed with copper to increase its strength and corrosion resistance, mainly to stress corrosion cracking [6]. Thus, copper-nickel alloys are used as marine piping materials and heat exchanger tubing materials, so corrosion behavior of such alloys in natural or industrial environments have been studied especially in long-term exposure times, e.g. for Cu90Ni10 (in at. %) alloy[12, 13].

Even though oxidation behavior of Ni, Cu and their alloys have been extensively studied, there is still a place for further investigations. Previously mentioned researchers investigated samples that were prepared in slightly different processes like sand-blasting or polishing ended on non-identical levels – the question arises if there is a relationship between the surface treatment and number of microstructural defects that were induced into subsurface area in preparation process. Over the years it was concluded that surface treatment processes may cause grain refinement and therefore increase the number of grain boundaries, what changes the alloy's mass gain during oxidation [14]. Also the number of defects in the surface region changes as energy is stored in the form of dislocations as a result of surface treatment Surface roughness is also influenced by surface preparation processes and therefore changes the oxidation kinetics – it has been recently proved, e.g. for ground superalloy IN718 [15], ground and polished DD6 Ni-base single crystal superalloy [16] and shot-peened nickel aluminum bronze [17]. However there is no much information in the literature about the influence of surface treatment on oxidation behavior of pure metals, what could have be beneficial in better understanding of such relation while avoiding the influence of alloying metals on oxidation behavior of the particular material. So far only few attempts had been performed, e.g. oxidation kinetics of shot-peened [18] and sandblasted pure titanium [19] have been investigated. Moreover it was already proved by Yuan et al [20] that increasing the time of sandblasting for copper samples improves samples' roughness and also their corrosion resistance. For nickel samples however it was concluded that after polishing performed in different conditions, the greater nickel roughness initially was, the greater values also the oxidation rate reached [21].

The aim of this paper is to determine the influence of different surface preparation on surface roughness and therefore on the oxidation kinetics and its mechanism at different temperatures in an ambient air for samples of high purity copper (99,99% Cu), nickel (99,99% Ni) and nickel-copper alloy (Ni30Cu70).

## 2. Material and methods

### 2.1. Surface preparation

From metal sheets rectangular samples with 17×11×2.5mm dimensions were cut. For each material 3 different surface preparation processes were chosen – initially each sample was ground using 220 grit SiC paper. First sample was left on this level of preparation (ground sample), second one, after grinding, was subsequently sand-blasted with alumina particles (sand-blasted sample) and the third one was further ground using SiC paper with increasing gradation (till 2400 grit) and then polished, finishing on 1 μm polishing SiO2 suspension (polished sample). Such prepared samples were ultrasonically cleaned with ethanol prior exposure.

### 2.2. Roughness evaluation

After surface preparation processes, surface roughness was measured with two different methods – conventional contact profilometer and using fractal analysis evaluation. Characterization of rough surfaces may be described in two-dimensional analysis by traditional statistical parameters like arithmetic mean deviation of the evaluated profile Ra or maximum height of the evaluated profile $R_z$ [22]. In case of three-dimensional analysis the above mentioned parameters are replaced by "S" letter. These methods of evaluation of surface topography and surface roughness are commonly used especially in controlling manufacturing processes like determination of surface roughness depended on hard turning process parameters [23] or description of surface topography after turning in the dry and MQCL (minimum quantity cooling lubrication) conditions [24].

Traditional method however is not perfect – it has been found that surfaces with the same parameter Ra may differ significantly [25]. Moreover above mentioned parameters, based on statistics depend on the sampling length and resolution of the apparatus [26].

Fractal geometry tends to overcome the limitations of dependency of conventional profilometers on sampling length and instrument's resolution by using scale-independent parameters describing self-similar objects [25]. Thanks to this concept, additional parameters like fractal dimension (FD) may describe how the object fills the space and therefore enable to quantitatively characterize topography and complexity of the surface [27]. This method has been recently successfully employed in determining surface roughness of nanostructured silver films [26,27] as well as in evaluation of oxidation behavior of Fe-Al-Cr-Zr-C alloys [28].

### 2.2.1. Conventional contact profilometer

Surface roughness is usually described by amplitude parameters, which enable to characterize the surface topography by measuring vertical characteristics difference. The most important parameters are: arithmetic average height (Ra) and maximum height of the profile called also as ten-point height (Rz). Calculation procedure is described in [22]. Measurements were perform using Hommel Werke T8000 profilometer: traverse length was set as 4mm and linear speed −0.5 mm/s. For each sample, 5 measurements of surface roughness were performed. Based on obtained data, the average value and standard deviation of measurements were calculated.

### 2.2.2. Fractal analysis

Surface roughness may be also described by fractal analysis. This is a mathematical method, which enable to describe self-similar objects. Surface roughness evaluation by fractal analysis starts with metallographic images of cross-sections of the samples prepared in different manners. Then image must be adjusted to meet the requirements of Sfrax 1.0 software [29] – microscopic photographs must be converted into binary scale images (metal – white and mounting – black color) and then digitalized into two-dimensional x-y data sets (the distance between subsequent points is constant). Such prepared files are the inputs to the Sfrax 1.0 software [30].

The outcome of data processing by this program contains information of the measured surface roughness profiles like following parameters: D – Fractal Dimension, LSFC – Length – Scale Fractal Complexity and $L_R$ – Relative length at given scale, that may be calculated by the following equation [31]:

$$L_R = \sum_i^N \frac{1}{\cos\theta_i} \frac{p_i}{L} \tag{1}$$

where: $L$ – total length, that is projected for all N virtual steps, $p_i$ – projected length for single virtual step, $\theta i$ – angle between normal and nominal surface.

Measurement of the apparent length of the given profile with decreasing rulers enables production of plots – relative length ($L_R$) vs. scale length ($r$). The slope of such curves in their linear part are used for calculation of other parameters describing surface roughness, that is Fractal Dimension D:

$$D = 1 + |slope| \tag{2}$$

and Length – Scale Fractal Complexity LSDC:

$$LSFC = 1000 \cdot (D - 1) \tag{3}$$

The more complex surfaces are, the higher values these parameters achieve [30].

## 2.3. Oxidation

Prior to oxidation process, the exposure temperature was established – pure copper samples were oxidized at the temperature range 650–750 °C, pure nickel was oxidized at 950–1050 °C, binary alloy samples – at 650–750 °C range. The ratio of chosen temperature to the melting point of particular sample was similar for different materials and so that the differences in the oxidation kinetics could be visible – it justifies the higher temperature range for nickel. Samples were then oxidized in a thermogravimetric furnace Xerion for 2 h each in the ambient air.

**2.4. Analysis and observations**

After exposure chosen samples were examined by using glow discharge optical emission spectrometry (GD-OES). Depth profiles were obtained by quantification described already by other authors [32–34].

Samples were then sputtered with very thin gold coating, so that electrolytically deposited layer of nickel had better adherence to the surface of the samples. The aim of nickel layer is to increase the contrast between the oxide layer and the resin during microscopic observations.

After electrolytical deposition, samples were cleaned and then mounted in epoxy resin. Microstructure of the samples was reveled thanks to grinding and polishing processes, that were finished on woven cloth with SiO2 suspension characterized by 0.25 μm grain size. Microstructure of the samples was then observed using scanning electron microscope Hitachi S3400N (SEM).

**3. Results**

**3.1. Surface roughness**

Roughness profiles of pure copper samples prepared in three different manners: by polishing, grinding and sand-blasting processes are presented in Fig. 1. The differences in profiles were so high, that the y-scale had to be adjusted – it proves that surface preparation influences greatly the surface roughness. The roughness profile was the highest for sand-blasted sample and definitively the lowest for polished sample. Similar observations were made for nickel and Ni30-Cu70 samples. Conclusions based on roughness profiles are proved by roughness parameters (Table 1).

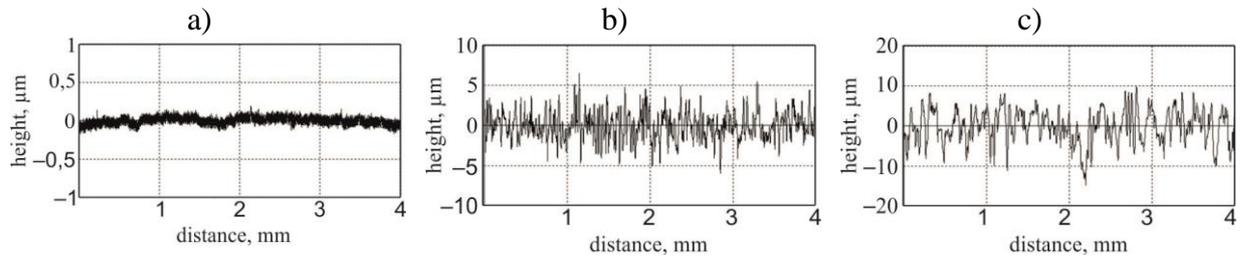

Fig. 1. Roughness profile as a function of measurement distance obtained for copper samples: a) polished, b) ground, c) sand-blasted.

Table 1. Results of roughness parameters obtained by conventional profilometer (P – polished sample, G – ground sample, SB – sand-blasted sample).

| Parameter | Copper | | | Ni30-Cu70 | | | Nickel | | |
|---|---|---|---|---|---|---|---|---|---|
| | P | G | SB | P | G | SB | P | G | SB |
| $R_a$, µm | 0.089 | 1.371 | 4.303 | 0.092 | 1.173 | 3.285 | 0.072 | 0.331 | 3.496 |
| SD | 0.006 | 0.062 | 0.425 | 0.001 | 0.068 | 0.356 | 0.001 | 0.021 | 0.296 |
| $R_z$, µm | 0.532 | 10.261 | 23.713 | 0.466 | 9.342 | 19.963 | 0.291 | 3.026 | 21.803 |
| SD | 0.072 | 0.732 | 3.341 | 0.052 | 0.865 | 1.300 | 0.144 | 0.367 | 1.288 |

To perform surface roughness evaluation by fractal analysis, binary scale images of metallographic cross-sections after preparation processes had to be make out. Examples of binary scale images of surface profile for copper samples are presented in Fig. 2. It is clearly seen, that for sand-blasted sample the surface is much more complex in comparison to polished and ground samples. Fig. 3 proves this observation and depicts plots of relative length vs. scale. For all of the materials sandblasted sample was characterized by the highest roughness and the polished one – by the lowest, although for Ni30Cu70, the difference between ground and polished sample was slightly visible. More information may be obtained by analyzing roughness parameters that were calculated by Sfrax 1.0 software (Table 2). For pure materials all the parameters increased in value in the following order: polished – the lowest values, sand-blasted – the highest. For nickel-copper alloy the roughness parameters obtained using fractal analysis are very similar, while, as observed for pure metals, sand-blasted sample exhibited the highest roughness values.

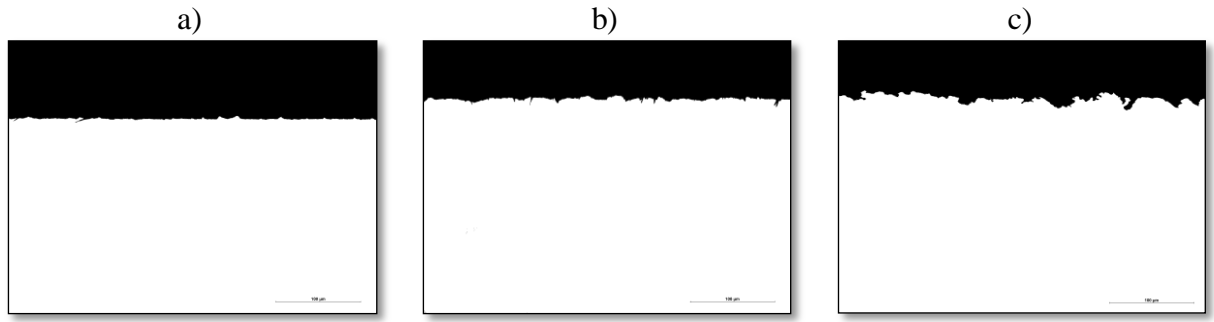

Fig. 2. Binary scale images of surface profile for Sfrax 1.0 software of copper samples: a) polished, b) ground, c) sand-blasted.

Table 2. Results of roughness parameters obtained by fractal analysis (P – polished sample, G – ground sample, SB – sand-blasted sample).

| Parameter | Copper | | | Ni30-Cu70 | | | Nickel | | |
|---|---|---|---|---|---|---|---|---|---|
| | P | G | SB | P | G | SB | P | G | SB |
| D | 1.016 | 1.084 | 1.105 | 1.016 | 1.015 | 1.100 | 1.004 | 1.036 | 1.11 |
| LSFC | 15.8 | 83.7 | 105.4 | 16.2 | 14.6 | 99.9 | 4.1 | 35.9 | 110.1 |
| $L_R$ 5 µm | 1.03 | 1.18 | 1.31 | 1.03 | 1.03 | 1.27 | 1.01 | 1.07 | 1.22 |

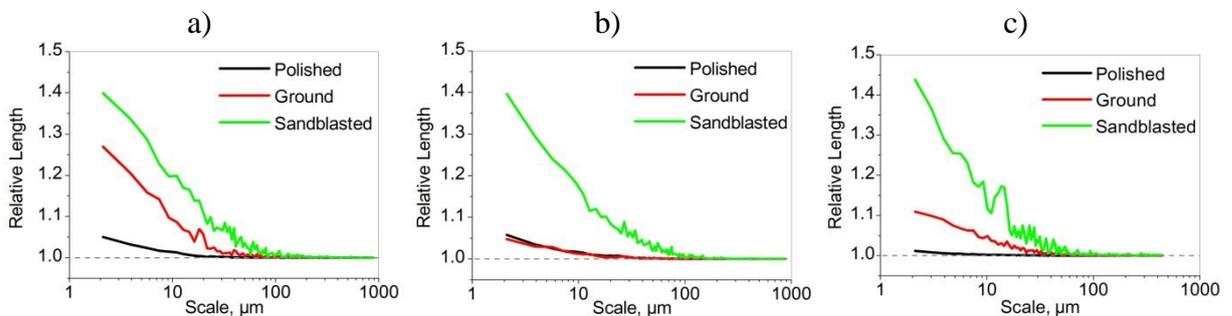

Fig. 3. Plots of relative length in function of scale obtained by Fractal analysis using Sfrax 1.0 software: a) copper, b) Ni30Cu70, c) nickel.

Nevertheless, surface roughness parameters obtained via profilometer and fractal analysis enabled to formulate similar conclusions – surface preparation process greatly affects surface roughness. Standard methods for quantification of material roughness, like $R_a$ and $R_z$ are useful for characterization of profile depth while fractal analysis gives more information about surface complexity and deformation. Combination of both methods are more and more widely used while describing oxidation processes [28,30].

**3.2. Oxidation of pure copper**

Normalized mass change as a function of oxidation time for copper samples is shown in Fig. 4. It is clearly seen that at 650 °C (Fig. 4a) and 700 °C (Fig. 4b), the mass gain of the particular samples was ordered in the same manner – the highest mass gain was obtained for polished samples, the lowest – for sand blasted ones. This order changed after exposure at 750 °C (Fig. 4c) – the highest mass gain was observed for sand-blasted sample – the mass gain increased almost three times in comparison to sand-blasted sample oxidized at 700 °C. The lowest mass gain was observed for polished sample – the value obtained was lower than that recorded for polished sample oxidized at 700 °C.

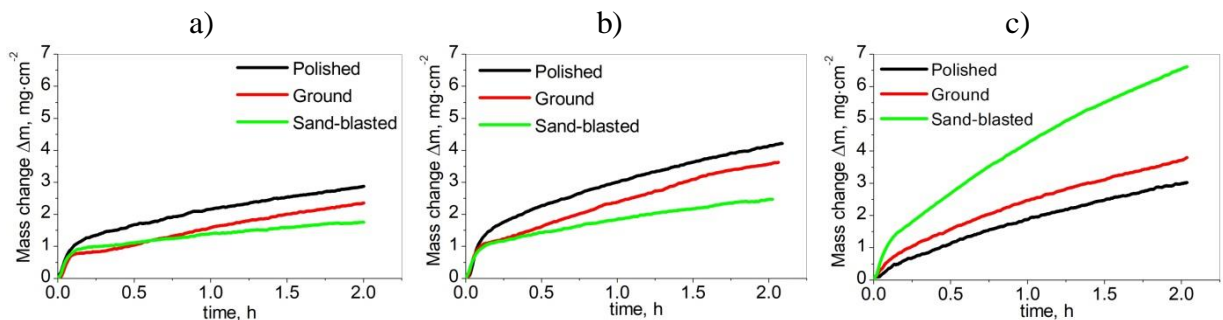

Fig. 4. Normalized mass change for pure copper oxidized for 2 h at a) 650°C, b) 700°C, c) 750°C.

Different surface preparation of the samples changes not only oxidation kinetics but also microstructure of the growing oxide layers. Observations of SEM images of cross-sections of the samples after oxidation (Fig. 5) present these differences. For polished (Fig. 5a, d and g) and ground (Fig. 5b, e and h) samples everything looks quite similar and relatively simple – thick layer of $Cu_2O$ is formed, which is covered by very thin layer of CuO oxide. Clear boundary between copper substrate and different oxides may be also proved by GD-OES depth profile (Fig. 6). This observation is in good agreement with literature – at 600–900 °C copper forms two oxide layers, but growth of $Cu_2O$ dominates [4].

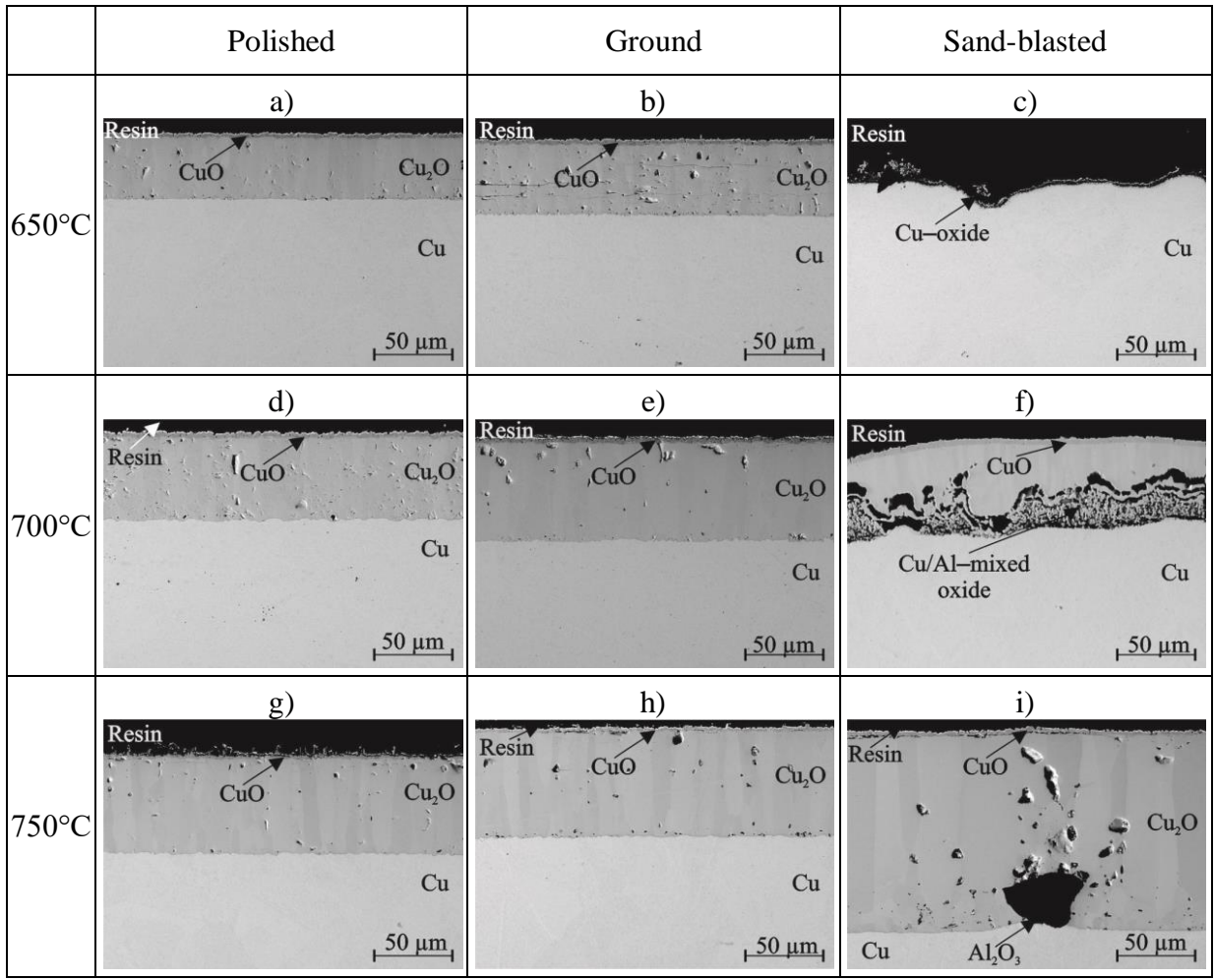

Fig. 5. SEM images of cross sections of copper samples oxidized at 650°C (a, b, c), 700°C (d, e, f), 750°C (g, h, i) and of samples characterized by different surface preparation: polished (a, d, g), ground (b, e, h) and sand-blasted (c, f, i).

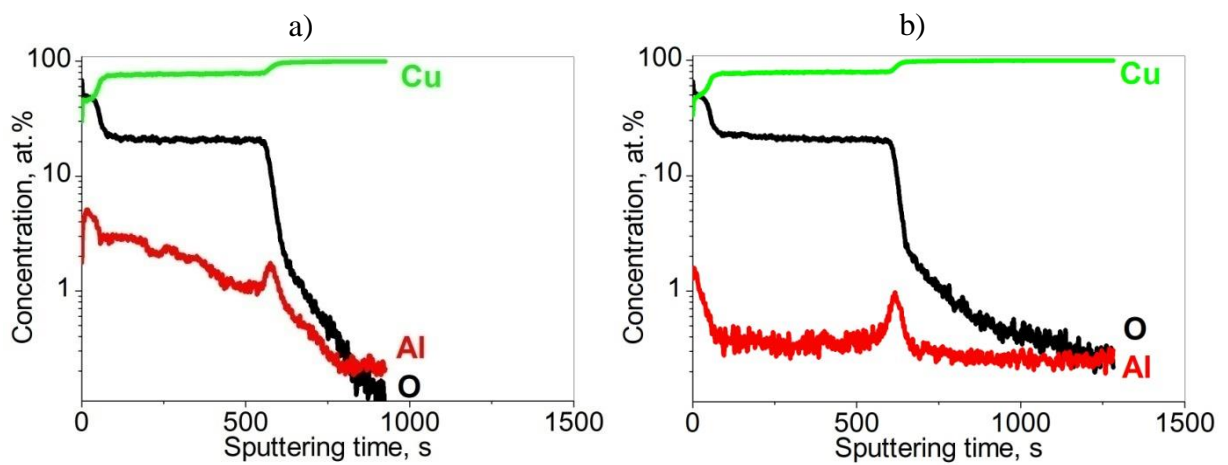

Fig. 6. The GD-OES depth profiles for a) polished and b) ground copper samples after exposure for 2 h at 650°C.

Situation is much more complicated for sand-blasted samples (Fig. 5c, f and i). Due to differences in coefficients of thermal conductivity of pure copper and its oxides, it was impossible to observe the microstructure of copper oxidized at 650 °C (Fig. 5c) – only small traces of copper oxide were noticed, because the rest of it spalled off while cooling to room temperature. However, the present traces of oxide look similar to the mixed Cu/Al oxide that was formed after exposure for 2 h at 700 °C (Fig. 5f). It may be assumed that mechanism of oxidation of these samples is similar – due to outward diffusion of copper, copper oxide is formed, but because of relatively slow rate of diffusion, copper oxide reacts with alumina that was induced by sand-blasting process and forms Cu/Al- mixed oxide (Fig. 7a). It is worth noticing, that oxide layer on these samples was possible to observe only in small areas of the samples, in comparison with rather even and compact oxide layers that were formed on polished and ground samples at these temperatures. It is a proof, that sand-blasting process results in decreasing oxide adhesion to the metallic substrate at lower temperatures.

It is clearly visible in the microphotograph of sand-blasted sample at the highest temperature –750 °C (Fig. 5i) that the oxidation mechanism changed – alumina particle does not form Al/Cu-mixed oxide, but is surrounded by columnar grains of $Cu_2O$. Moreover its position inside the oxide layer is a proof of outward diffusion of copper. However, remembering how the surface roughness looked like before oxidation (Fig. 2c) it is interesting to observe, how smooth the boundary between oxide and substrate looks like – in this case probably the diffusion rate increased by including not only outward diffusion of copper but also inward diffusion of oxygen in the initial point of oxidation. It also prevented creation of mixed Al/Cu oxide (Fig. 7b).

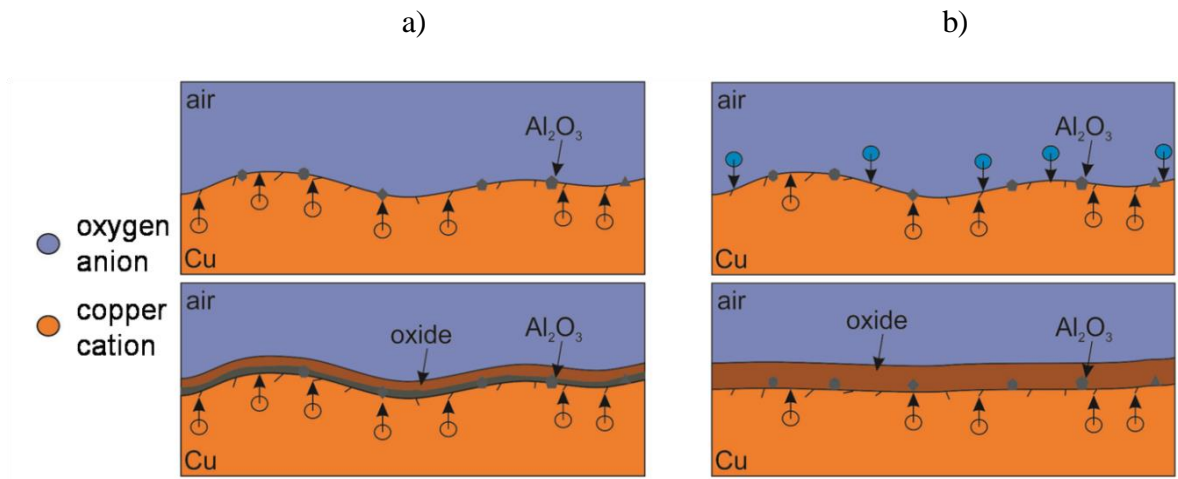

Fig. 7. Proposed mechanisms of oxidation of sand-blasted samples at a) 700°C and b) 750°C.

Above formulated theory about the influence of diffusion on kinetics of oxidation can be further justified by calculating the ratio of $Cu_2O$ to CuO oxides thicknesses (table 3). It is already known, that in case of oxidation under the same conditions (temperature and oxygen partial pressure), ratio of particular phases thickness of multilayer scale depends on diffusion of reagents in the growing oxide so that the oxidation kinetics is also determined by the diffusion rate and in consequence – by concentration of lattice defects [35]. Analyzing table 3 one can observe that at 650 °C ratio of thickness of $Cu_2O$ to CuO is almost similar for samples prepared in polishing and grinding processes. Situation changes at 700 °C – for polished sample ratio of $Cu_2O$ to CuO equals 17.18. This value increases further for ground sample to 20.20. Unexpectedly mentioned ratio decreased for sand-blasted sample to 14.76, but microphotograph of this sample (Fig. 5f) has already reveled that Al/Cu-mixed oxide formed which considerably slowed the oxidation process. Such effect is not observed after exposure at 750 °C. For this samples, the lowest value of Cu2O to CuO thicknesses ratio was observed for polished sample (26.20) and slightly greater for ground sample (29.64). For sand-blasted sample however, the value of measured ratio was almost two times higher than for polished sample. It proves, that the diffusion process is connected with increased number of defects occurred for this sample and increased the oxidation kinetics.

Table 3. Measured thicknesses of two formed copper oxides and the relationship between them.

| Temp. | Sample preparation | CuO | | Cu$_2$O | | Cu$_2$O to CuO thickness ratio |
|---|---|---|---|---|---|---|
| | | Average thickness, µm | Standard deviation | Average thickness, µm | Standard deviation | |
| 650°C | Polished | 2.48 | 0.81 | 39.34 | 0.55 | 15.87 |
| | Ground | 2.95 | 0.54 | 45.48 | 0.85 | 15.40 |
| | Sand-blasted | - | - | - | - | - |
| 700°C | Polished | 3.10 | 0.76 | 53.29 | 0.96 | 17.18 |
| | Ground | 3.11 | 0.51 | 62.85 | 1.23 | 20.20 |
| | Sand-blasted | 3.66 | 0.40 | 54.03 | 4.21 | 14.76 |
| 750°C | Polished | 2.35 | 0.75 | 61.52 | 0.87 | 26.20 |
| | Ground | 2.29 | 0.46 | 67.83 | 0.75 | 29.64 |
| | Sand-blasted | 2.89 | 0.80 | 125.53 | 0.78 | 43.45 |

## 3.3. Oxidation of pure nickel

Normalized mass change for pure nickel samples oxidized at 3 different temperature values is shown in Fig. 8. Also in this case it is seen, that the mass gain is influenced by surface preparation, but it is worth noticing, that the mass gain obtained for these samples does not vary significantly – even though at 950 and 1000 °C (Fig. 8a and b) the order of the samples characterized by particular surface preparation process is different during the exposure time – the final mass gain for these samples is almost equal. The change is observed for last samples, oxidized at 1050 °C – the sand blasted sample was characterized by the highest mass gain, and ground sample – the lowest, even though its oxidation rate at the first 0.5 h was the fastest.

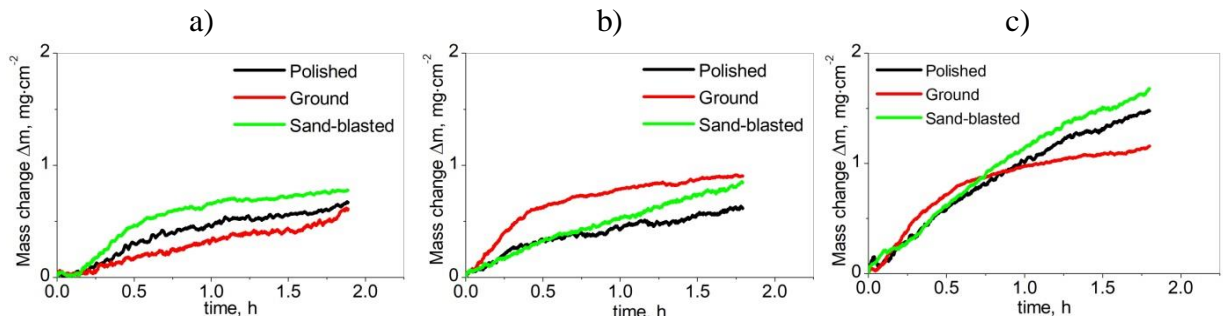

Fig. 8. Normalized mass change for pure nickel oxidized for 2 h at a) 950°C, b) 1000°C, c) 1050°C.

Similarly to the copper samples prepared before oxidation in grinding and polishing processes, the oxide layer that forms on its surface does not differ significantly – the NiO layers on polished (Fig. 9a, d and g) and ground (Fig. 9b, e and h) samples is compact and demonstrates good adherence to the nickel substrate. Again the situation changes for sand-blasted samples, because their surfaces are rougher and also contaminated by alumina particles that were inducted into the surface in preparation process. Microphotographs of these samples (Fig. 9c, f and i) show that alumina particles inhibits the formation on nickel oxide – in Fig. 9c, thickness of NiO that grows on $Al_2O_3$ is lower than the same oxide growing directly on nickel substrate. The explanation is simple - NiO is formed by inward diffusion of oxygen anions, so in case of alumina particles, the diffusing anions have to overcome longer way. In Fig. 9i, alumina particle is so high, that the growing NiO does not overgrow its surface as in Fig. 9c and the discontinuity in NiO-layer is observed. Taking that into account one can conclude that sand-blasting process influences the adhesion between nickel surface and growing oxide.

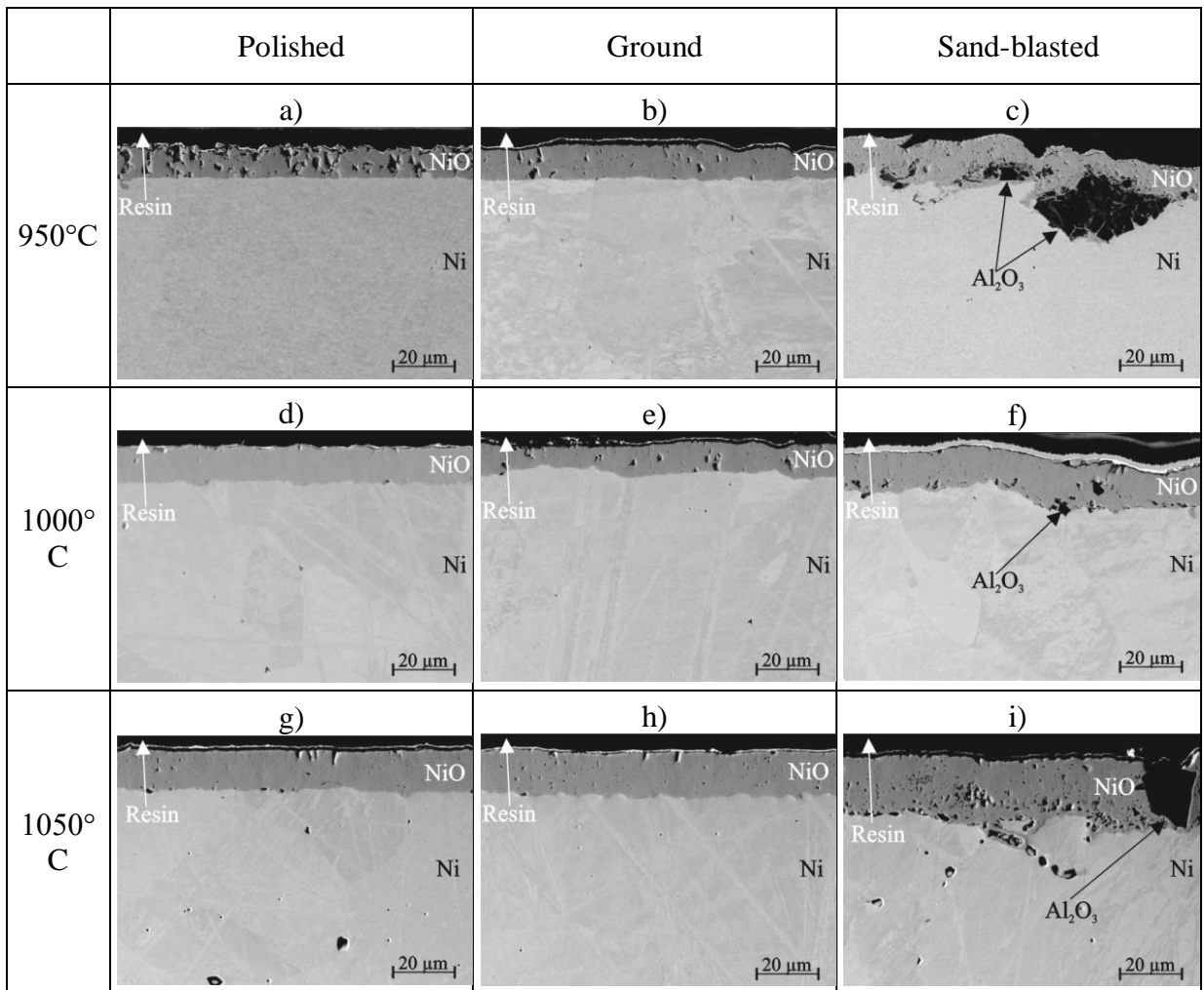

Fig. 9. SEM images of cross sections of nickel samples oxidized at 650°C (a, b, c), 700°C (d, e, f), 750°C (g, h, i) and of samples characterized by different surface preparation: polished (a, d, g), ground (b, e, h) and sand-blasted (c, f, i).

How the inducted alumina particles change the chemical composition of oxide layer is clearly visible on GD-OES depth profiles (Fig. 10). For polished and ground samples (Fig. 10a and b) only NiO is formed – there are only small, irrelevant amounts of aluminum. In Fig. 10c however great enrichment of aluminum can be observed. Outer layer contains lower amount of this element – it proves, that due to the diffusion of oxygen anions, the alumina particles are eventually covered by growing NiO. Aluminum is visible not only in the oxide but also in the substrate. It means that during sand-blasting process alumina particles gain great kinetic energy so that it is possible for them to be inserted in relatively deep regions of the substrate.

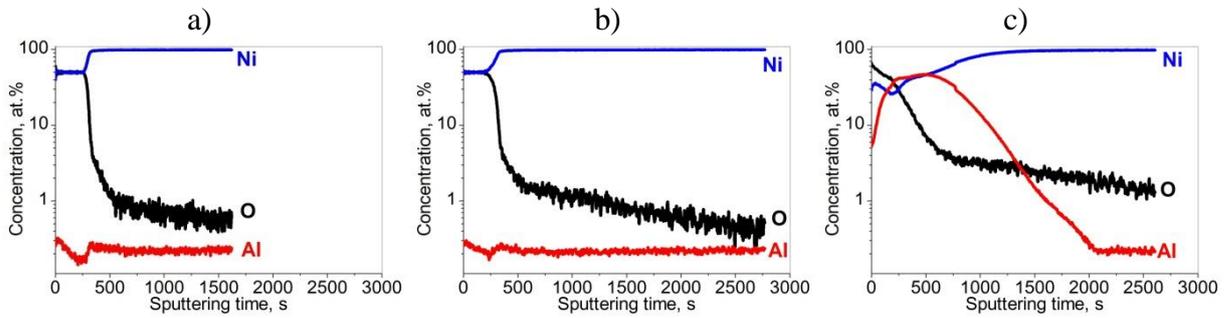

Fig. 10. The GD-OES depth profiles for a) polished, b) ground and c) sand-blasted nickel samples after exposure for 2 h at 950°C.

### 3.4. Nickel-copper alloy oxidation

Plots of mass change per unit area observed for Ni30Cu70 samples oxidized for 2 h at different temperature are depicted in Fig. 11. It is clearly seen that oxidation kinetics changes as a result of surface preparation – regardless the temperature changes, the lowest mass gain was revealed sand-blasted samples. At lower temperatures (Fig. 11a and b) ground samples obtained the greatest mass change, whereas at 750 °C, the polished sample started to prevail. The highest mass gain for all the samples was obtained after exposure at 700 °C (Fig. 11b), but the values were significantly lower than that observed for oxidation of pure copper at the same temperature (Fig. 4b).

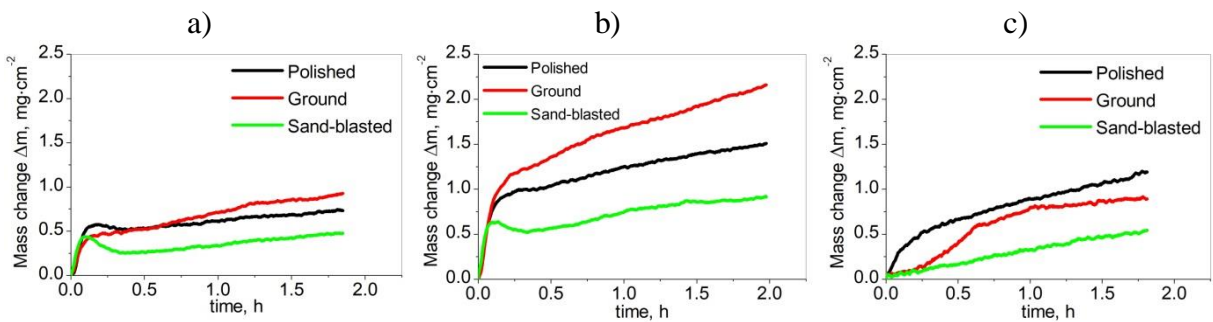

Fig. 11. Normalized mass change for nickel-copper alloy (Ni30Cu70) oxidized for 2 h at a) 650°C, b) 700°C, c) 750°C.

SEM images of cross sections of samples after exposure at different temperatures help to explain the effect observed on kinetics plots. For polished samples (Fig. 12a, d and g) the oxide layer that grows on Ni30-Cu70 substrate looks similar: CuO forms the outer layer beneath which mixed nickel/copper oxide is situated. Difference is observed in their thickness – the higher the temperature is, the thicker both of these layers are. As a result of outward diffusion of copper, outer surface of Ni30-Cu70 substrate is depleted in this element. GD-OES depth profile after exposure at 750 °C (Fig. 13a) outer layer is created almost completely by CuO – small peak of nickel indicates electrodeposited layer of this element on the sample after oxidation exposure. Below CuO enrichment in nickel and depletion in copper are observed – this part of the plot presents formation of Ni/Cu mixed oxide.

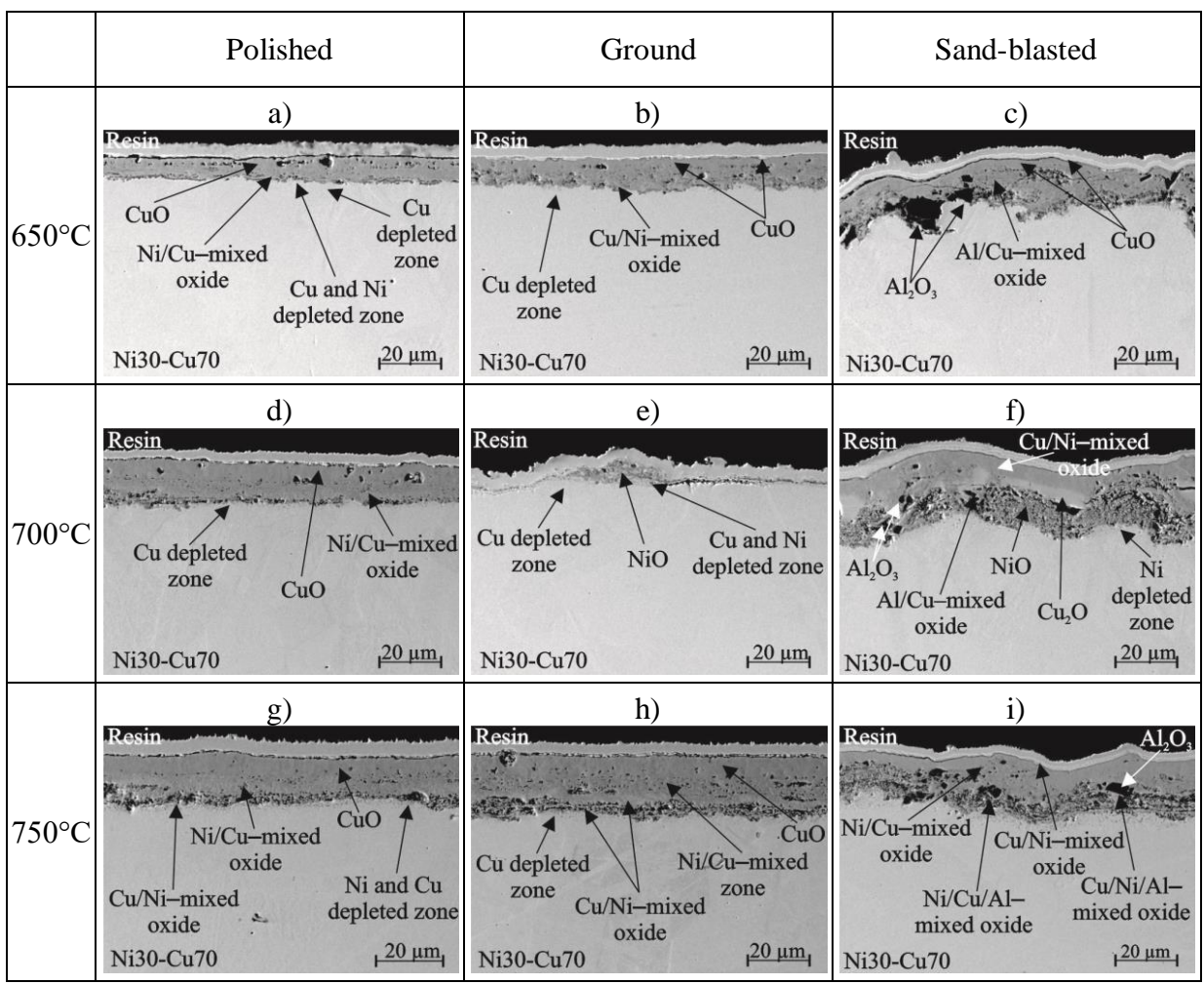

Fig. 12. SEM images of cross sections of Ni30-Cu70 samples oxidized at 650°C (a, b, c), 700°C (d, e, f), 750°C (g, h, i) and of samples characterized by different surface preparation: polished (a, d, g), ground (b, e, h) and sand-blasted (c, f, i). In situations when mixed oxide is formed, the former element indicates the element with highest content in the oxide while the latter –the lowest.

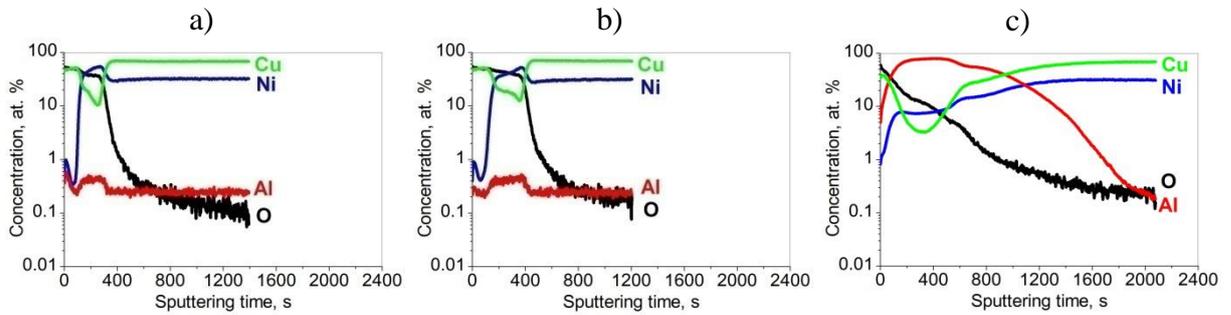

Fig. 13. The GD-OES depth profiles for a) polished, b) ground and c) sand-blasted Ni30-Cu70 samples after exposure for 2 h at 750°C.

Quite similar observations can be made for ground samples (Figs. 12b, e and h, 13b). Outer layer is also formed by CuO and inner – by mixed Cu/Ni oxide. It is impossible to observe the oxide layer that grew on ground sample that was oxidized at 700 °C – oxidation products spalled off during cooling to the room temperature. Because this sample exhibited the highest mass gain of all the samples (Fig. 11b) it may be assumed that the oxide layer on this sample reached its critical thickness what caused the effect of flaking off.

It is once again visible that sand-blasting process changes the oxidation mechanism the most. For the samples prepared in this way (Fig. 12c, f and i) alumina particles were observed in two forms – as a separate particles that were stuck in the substrate and were then surrounded by growing oxide and as a component of mixed oxides that also appeared during oxidation.

## 5. Discussion

In this paper three different materials: high purity copper, high purity nickel and high purity binary nickel-copper (Ni30-Cu70) alloy were treated in three different preparation processes: grinding, polishing and sand-blasting and then oxidized at three different temperature values. Two methods of evaluation of surface roughness have been applied – they both confirmed that surface preparation process influences greatly the surface roughness. As it was predicted, the roughest surface was obtained for sand-blasted samples and the smoothest – for polished ones. The parameters of roughness obtained by conventional contact profilometer the values changed almost of one order of magnitude (table 1). It was also noticed, that even though samples of different materials in particular processes were prepared in exactly the same way, roughness parameters varied for each material – for example ground sample of nickel had almost 4 times lower value of $R_a$ than copper and Ni30-Cu70 samples (table 1). The relationship between grinding conditions and surface roughness was already studied elsewhere [36,37].

Surface preparation processes do not only influence surface roughness – they may also induce an increase in residual stresses and number of defects in the subsurface area and therefore – to change the oxidation mechanism by promoting fast diffusion paths [38,39]. Putting together increased area exposed to oxidant and faster diffusion of components, one can assume that the more convoluted the surface is due to preparation process, the faster the oxidation proceeds. Such theory has been already proposed by Nowak et al. [40] for Ni-based superalloy IN 713C samples prepared in grinding and polishing processes. The mentioned theory seems to be valid also for oxidation of pure copper at 750 °C (Fig. 4c) and nickel at 950 and 1050 °C (Fig. 8) but fails in other cases – for example for Ni30Cu70 sand-blasted samples obtained the lowest mass gain at each temperature (Fig. 11). It is known however that oxidation mechanisms for pure metals and commercially available alloys vary significantly, so that further investigations about the induction of defects in the subsurface area in preparation processes should be performed. Moreover SEM analyses and GD-OES depth profiles revealed that intrusions of aluminum may be responsible for changes in oxidation kinetics – it is widely know that any impurities have great influence on oxidation process [2]. Considering the results obtained in this research it is impossible to distinguish the influence of surface roughness, number of defects and addition of $Al_2O_3$ particles on oxidation kinetics.

From the other side comparison of ground and polished samples, which theoretically did not contain any contaminations before oxidation process, also does not give a straight confirmation of the theory. For oxidation of pure copper polished sample presented the highest mass gain at 650 and 700 °C, but the lowest at 750 °C (Fig. 4). In the contrary during oxidation of nickel-copper alloy, ground samples had the highest mass gain at lower temperatures (Fig. 11) but polished sample won the competition at 750 °C. In case of pure copper the explanation may be as followed: copper oxide grows due to the outward diffusion of metal cations, so in case of smooth surface all of copper ions may connect with oxide anions. The more convoluted surface however, for example, in concave areas, the number of oxide anions may be to scarce to connect all of copper cations, and thus for flat, polished samples, the oxidation rate was higher. In case of oxidation of nickelcopper alloy, formation of mixed oxide may play a key role which changes the oxidation mechanism observed for pure nickel and copper separately. The exact explanation is not yet known.

It is visible that there is a relationship between surface treatment and oxidation rate and mechanism for pure metals and binary alloy, but to confirm the mentioned theory and form more specific conclusions another testing must be performed, like measurement of residual stresses of samples before oxidation and oxidation tests with application of markers. Both investigation techniques will be used in further research.

## 6. Conclusions:

In the present work, the effect of surface deformation on high temperature oxidation of Ni, Cu and Ni-Cu alloy has been investigated. Based on obtained results of research following conclusions are formulated:

1. Surface roughness parameters $R_a$ and $R_z$ changes about one order of magnitude between different preparation processes: the highest average value of $R_a$ was measured for sand-blasted copper sample ($R_a$=4.303), ground copper sample ($R_a$=1.371) and polished nickel sample ($R_a$=0.072). Also fractal analysis results prove the differences in surface roughness after polishing, grinding and sandblasting processes. Moreover the latter results in introduction of alumina particles in subsurface area.

2. Alumina particles sticking in subsurface area cause formation of Cu/Al-mixed oxide for copper sample oxidized at 700°C and for all sand-blasted samples of Ni-Cu alloy. $Al_2O_3$ is also observed as a separate particles surrounded by the growing oxide as in oxidation of pure nickel, which may lead to discontinuity in NiO layer and lower the oxide adhesion to the metal surface.

3. Morphology of the polished and ground samples does not vary significantly for pure Cu, Ni and binary Ni-Cu alloy after oxidation at investigated temperatures, although differences in oxide layer thicknesses and mass gains were observed –for pure nickel highest mass gain was observed for ground sample at 1000°C ($\Delta m = 0.904$ mg·cm$^{-2}$), while the mass gain for polished sample was higher at 950 and 1050°C ($\Delta m = 0.670$ and 1.474 mg·cm$^{-2}$, respectively). Polished copper sample obtained greater mass gain that ground copper sample at 650 and 700°C ($\Delta m = 2.874$ and 4.214 mg·cm$^{-2}$, respectively), but for oxidation at 750° the order changed – ground sample obtained greater mass gain ($\Delta m = 3.801$ mg·cm$^{-2}$). For Ni-Cu alloy samples behaved exactly in the opposite way – at 650 and 700°C ground sample obtained higher mass gain ($\Delta m = 0.927$ and 2.381 mg·cm$^{-2}$, respectively) while at 750°C it was polished sample ($\Delta m = 1.923$ mg·cm$^{-2}$).

4. Observation of pure copper sample revealed that initially rough surface after sand-blasting process smoothed after exposure at 750°C. In connection with great increase in oxidation kinetics ($\Delta m$ after 2 h of oxidation equaled 3.020; 3.795 and 6.615 mg·cm$^{-2}$ for polished, ground and sand-blasted samples, respectively) and differences in $Cu_2O$ to CuO thickness ratio (26,20; 29,64; 43,45 for polished, ground and sand-blasted samples, respectively) it proves that additional diffusion mechanism has occurred for this sample at high-temperature due to increased number of defects introduced in sand-blasting process.

5. It has been proved that surface preparation has influence on oxidation kinetics of Ni, Cu and Ni30Cu70 alloy. However observed differences in kinetics of the oxidation and morphology of the oxides depend not only on surface preparation but also on the nature of base materials and their oxidation mechanism.

**7. Funding**

This project is financed within the Marie Curie COFUND scheme and POLONEZ program from the National Science Centre, Poland, POLONEZ Grant No. 2015/19/P/ST8/03995. This project has received funding from the European Union's Horizon 2020 Research and Innovation Programme under the Marie Skłodowska Curie Grant Agreement No. 665778.

**8. Reference List**

[1] P. Kofstad, High Temperature Corrosion, Elsevier Applied Science, London/New York, 1988.

[2] D.J. Young, High Temperature Oxidation and Corrosion of Metals, Elsevier Science, 2016. https://books.google.pl/books?id=TVXBBwAAQBAJ.

[3] S. Mrowec, A. Stokłosa, Oxidation of copper at high temperatures, Oxid. Met. 3 (1971) 291–311, https://doi.org/10.1007/BF00603530.

[4] Y. Zhu, K. Mimura, M. Isshiki, Oxidation mechanism of copper at 623–1073 K, Mater. Trans. – MATER. TRANS. 43 (2002) 2173–2176, https://doi.org/10.2320/matertrans.43.2173.

[5] K. Mimura, L. Jae Won, M. Isshiki, Y. Zhu, Q. Jiang, Brief review of oxidation kinetics of copper at 350 °C to 1050 °C, Metall. Mater. Trans. A (2006), https://doi.org/10.1007/s11661-006-1074-y.

[6] R.N. Caron, Copper Alloys: Properties and Applications; 2001. doi:10.1016/B0-08-043152-6/00292-8.

[7] V.V. Prisedsky, V.M. Vinogradov, Fragmentation of diffusion zone in high-temperature oxidation of copper, J. Solid State Chem. 177 (2004) 4258–4268, https://doi.org/10.1016/j.jssc.2004.07.058.

[8] E.A. Goldstein, T.M. Gür, R.E. Mitchell, Modeling defect transport during Cu oxidation, Corros. Sci. 99 (2015) 53–65, https://doi.org/10.1016/j.corsci.2015.05.067.

[9] R. Haugsrud, On the high-temperature oxidation of nickel, Corros. Sci. 45 (2003) 211–235, https://doi.org/10.1016/S0010-938X(02)00085-9.

[10] L.Z. Mohamed, W.A. Ghanem, O.A. El Kady, M.M. Lotfy, H.A. Ahmed, F.A. Elrefaie, Oxidation characteristics of porous-nickel prepared by powder metallurgy and castnickel at 1273K in air for total oxidation time of 100h, J. Adv. Res. 8 (2017) 717–729, https://doi.org/10.1016/j.jare.2017.08.004.

[11] T. Yonezawa, Nickel Alloys: Properties and Characteristics, 2012. doi:10.1016/ B978-0-08-056033-5.00016-1.

[12] A.L. Ma, S.L. Jiang, Y.G. Zheng, W. Ke, Corrosion product film formed on the 90/10 copper–nickel tube in natural seawater: composition/structure and formation mechanism, Corros. Sci. 91 (2015) 245–261, https://doi.org/10.1016/j.corsci.2014.11.028.

[13] R.E. Melchers, Bi-modal trends in the long-term corrosion of copper and high copper alloys, Corros. Sci. 95 (2015) 51–61, https://doi.org/10.1016/j.corsci.2015.02.001.

[14] C. Sudbrack, D.L. Beckett, R.A. Mackay, Effect of surface preparation on the 815°C oxidation of single-crystal nickel-based superalloys, JOM (2015), https://doi.org/10.1007/s11837-015-1639-6.

[15] D. Pradhan, G.S. Mahobia, K. Chattopadhyay, V. Singh, Effect of surface roughness on corrosion behavior of the superalloy IN718 in simulated marine environment, J. Alloys Compd. 740 (2018) 250–263, https://doi.org/10.1016/j.jallcom.2018.01.042.

[16] H. Pei, Z. Wen, Z. Li, Y. Zhang, Z. Yue, Influence of surface roughness on the oxidation behavior of a Ni-4.0Cr-5.7Al single crystal superalloy, Appl. Surf. Sci. 440 (2018) 790–803, https://doi.org/10.1016/j.apsusc.2018.01.226.

[17] C. Wang, C. Jiang, Z. Chai, M. Chen, L. Wang, V. Ji, Estimation of microstructure and corrosion properties of peened nickel aluminum bronze, Surf. Coatings Technol. 313 (2017) 136–142, https://doi.org/10.1016/j.surfcoat.2017.01.073.

[18] A. Kanjer, V. Optasanu, M.C.M. de Lucas, O. Heintz, N. Geoffroy, M. François, P. Berger, T. Montesin, L. Lavisse, Improving the high temperature oxidation resistance of pure titanium by shot-peening treatments, Surf. Coatings Technol. 343 (2018) 93–100, https://doi.org/10.1016/j.surfcoat.2017.10.065.

[19] X.P. Jiang, X.Y. Wang, J.X. Li, D.Y. Li, C.-S. Man, M.J. Shepard, T. Zhai, Enhancement of fatigue and corrosion properties of pure Ti by sandblasting, Mater. Sci. Eng. A 429 (2006) 30–35, https://doi.org/10.1016/j.msea.2006.04.024.

[20] L. Yuan, X. Chen, S. Maganty, J. Cho, C. Ke, G. Zhou, Enhancing the oxidation resistance of copper by using sandblasted copper surfaces, Appl. Surf. Sci. 357 (2015) 2160–2168, https://doi.org/10.1016/j.apsusc.2015.09.203.

[21] A.M. Huntz, B. Lefevre, F. Cassino, Roughness and oxidation: application to NiO growth on Ni at 800°C, Mater. Sci. Eng. A 290 (2000) 190–197, https://doi.org/10.1016/S0921-5093(00)00944-8.

[22] E.S. Gadelmawla, M.M. Koura, T.M.A. Maksoud, I.M. Elewa, H.H. Soliman, Roughness parameters, J. Mater. Process. Technol. 123 (2002) 133–145, https://doi.org/10.1016/S0924-0136(02)00060-2.

[23] P. Nieslony, G.M. Krolczyk, S. Wojciechowski, R. Chudy, K. Zak, R.W. Maruda, Surface quality and topographic inspection of variable compliance part after precise turning, Appl. Surf. Sci. 434 (2018) 91–101, https://doi.org/10.1016/j.apsusc.2017.10.158.

[24] G.M. Krolczyk, R.W. Maruda, J.B. Krolczyk, P. Nieslony, S. Wojciechowski, S. Legutko, Parametric and nonparametric description of the surface topography in the dry and MQCL cutting conditions, Measurement 121 (2018) 225–239, https://doi.org/10.1016/j.measurement.2018.02.052.

[25] Z. Chen, Y. Liu, P. Zhou, A comparative study of fractal dimension calculation methods for rough surface profiles, Chaos, Solitons Fract. 112 (2018) 24–30, https://doi.org/10.1016/j.chaos.2018.04.027.

[26] M. Nasehnejad, G. Nabiyouni, M.G. Shahraki, Fractal analysis of nanostructured silver film surface, Chin. J. Phys. 55 (2017) 2484–2490, https://doi.org/10.1016/j.cjph.2017.10.015.

[27] G. Nabiyouni, M. Nasehnejad, Conventional and fractal analyses and nanoscale behavior studies of electrodeposited silver films, Phys. B Condens. Matter. 548 (2018) 46–52, https://doi.org/10.1016/j.physb.2018.08.014.

[28] V. Hotař, A. Hotař, Fractal dimension used for evaluation of oxidation behaviour of Fe-Al-Cr-Zr-C alloys, Corros. Sci. 133 (2018) 141–149, https://doi.org/10.1016/j.corsci.2018.01.017.

[29] Sfrax 1.0., (n.d.). http://www.surfract.com/sfrax.html (accessed September 24, 2018).

[30] W. Nowak, D. Naumenko, G. Mor, F. Mor, D.E. Mack, R. Vassen, L. Singheiser, W.J. Quadakkers, Effect of processing parameters on MCrAlY bondcoat roughness and lifetime of APS–TBC systems, Surf. Coatings Technol. 260 (2014) 82–89, https://doi.org/10.1016/j.surfcoat.2014.06.075.

[31] C.A. Brown, W.A. Johnsen, R.M. Butland, J. Bryan, Scale-sensitive fractal analysis of turned surfaces, CIRP Ann. 45 (1996) 515–518, https://doi.org/10.1016/S0007-8506(07)63114-X.

[32] J.P. Pfeifer, H. Holzbrecher, W. Quadakkers, U. Breuer, W. Speier, Quantitative analysis of oxide films on ODS-alloys using MCs+-SIMS and e-beam SNMS, Fresenius J. Anal. Chem. (1993), https://doi.org/10.1007/BF00321410.

[33] W.J. Quadakkers, A. Elschner, W. Speier, H. Nickel, Composition and growth mechanisms of alumina scales on FeCrAl-based alloys determined by SNMS, Appl. Surf.Sci. 52 (1991) 271–287, https://doi.org/10.1016/0169-4332(91)90069-V.

[34] W.J. Nowak, Characterization of oxidized Ni-based superalloys by GD-OES, J. Anal. At. Spectrom. 32 (2017) 1730–1738, https://doi.org/10.1039/C7JA00069C.

[35] S. Mrowec, Podstawy teorii utleniania metali i stopów, Wydawnictwa Naukowo-Techniczne, Warszawa, 1964.

[36] R.L. Hecker, S.Y. Liang, Predictive modeling of surface roughness in grinding, Int. J. Mach. Tools Manuf. 43 (2003) 755–761, https://doi.org/10.1016/S0890-6955(03)00055-5.

[37] P. Puerto, R. Fernández, J. Madariaga, J. Arana, I. Gallego, Evolution of surface

roughness in grinding and its relationship with the dressing parameters and the radial wear, Proc. Eng. 63 (2013) 174–182, https://doi.org/10.1016/j.proeng.2013.08.181.

[38] L. Yang, M. Chen, Y. Cheng, J. Wang, L. Liu, S. Zhu, F. Wang, Effects of surface finish of single crystal superalloy substrate on cyclic thermal oxidation of its nanocrystalline coating, Corros. Sci. 111 (2016) 313–324, https://doi.org/10.1016/j.corsci.2016.04.023.

[39] L. Cooper, S. Benhaddad, A. Wood, D.G. Ivey, The effect of surface treatment on the oxidation of ferritic stainless steels used for solid oxide fuel cell interconnects, J. Power Sources 184 (2008) 220–228, https://doi.org/10.1016/j.jpowsour.2008.06.010.

[40] W.J. Nowak, B. Wierzba, Effect of surface treatment on high-temperature oxidation behavior of IN 713C, J. Mater. Eng. Perform. 27 (2018) 5280–5290, https://doi.org/10.1007/s11665-018-3621-2.

.